\documentclass[11pt,a4paper]{article}
\usepackage[utf8]{inputenc}
\usepackage{amsmath}
\usepackage{amsfonts}
\usepackage{amssymb}
\usepackage{graphicx}
\usepackage{enumerate}
\usepackage{xfrac}
\usepackage{textcomp}
\usepackage{color}   
\usepackage{hyperref}
\hypersetup{
    colorlinks=true, 
    linktoc=all,     
    linkcolor=black,  
}

\usepackage{IEEEtrantools}
\usepackage[left=2.3cm,right=2cm,top=3cm,bottom=2cm]{geometry}
\usepackage{authblk}
\usepackage{geometry}
\usepackage{graphicx}
\usepackage{dcolumn}
\usepackage{bm}
\usepackage{mathtools}
\usepackage{amsmath}
\usepackage{amsfonts}
\usepackage{amssymb}
\usepackage{enumerate}
\usepackage{mathtools}
\usepackage{float}
\usepackage{xfrac}
\usepackage{hyperref}
\usepackage{tabularx}
\usepackage{xcolor}
\hypersetup{
    colorlinks=false, 
    linktoc=all,     
    linkcolor=black,  
    citecolor=black,
}

\usepackage[numbers,sort&compress]{natbib}

\usepackage[font={small}]{caption}

\title{\textbf{Optothermal Revolution: Colloids in an Optical Ring Trap}}
\author[$\dagger$,$*$]{Rahul Chand}
\author[$\dagger$]{Ashutosh Shukla}
\author[$\dagger$,$*$]{G V Pavan Kumar}

\affil[$\dagger$]{Department of Physics, Indian Institute of Science Education and Research (IISER) Pune, Pune, 411008, India}

\affil[$*$]{email: \textrm{rahul.chand@students.iiserpune.ac.in, pavan@iiserpune.ac.in }}
\date{}
\newcommand{\beginsupplement}{%
        \setcounter{table}{0}
        \renewcommand{\thetable}{S\arabic{table}}%
        \setcounter{figure}{0}
        \renewcommand{\thefigure}{{\arabic{figure}}}%
     }
     
\begin{document}
\beginsupplement
\maketitle
\renewcommand{\figurename}{{Figure}}
\renewcommand\thesection{S{\arabic{section}}}



\begin{abstract}
Directional motion is commonly observed in various living active systems, such as bacterial colonies moving through confined environments. In these systems, the dynamics arise from the collective effects of mutual interactions between individual elements, as well as their interactions with obstacles or boundaries. In this study, we turn our focus to an artificial system and experimentally investigate the emergence of directional revolution in dimer and trimer structures composed of colloidal particles in ring-shaped optical illumination. In this case, the movement of these colloidal structures is exclusively facilitated by optothermal interactions—without any direct mechanical force applied from external optical field. Depending on the optical absorption properties of the colloidal particles, these optothermal interactions can exhibit both attractive and repulsive characteristics. The attractive interactions provide the necessary driving force that propels the motion, while the repulsive interactions serve to control the structural parameters of the system. The arrangement and interaction of the colloidal particles within these dimer and trimer structures fuel the controlled, directional revolution, with the optical gradient force acting as a confining factor, guiding the movement along a specific path. Notably, the dynamics of these systems can be tuned by altering the intensity of the optical field. This study can be useful as a model for understanding insights into biological systems where group dynamics and environmental interactions are key to coordinated movement.
\end{abstract}
\section*{Introduction}
Active motion refers to the self-propelled movement of particles, organisms, or agents driven by an internal energy source \cite{hierarchically_assemble_active, thermodynamics_active_matter, hydrodynamic_active_matter, computation_active_matter}. At all scales, the active systems operate far from thermodynamic equilibrium. Unlike passive systems, which rely on external forces, active matter can move and maintain motion by consuming energy from its surroundings or from internal metabolic processes. 

These active systems can exhibit various collective states such as group formation, collective motility, and dynamic pattern formations \cite{vicsek2012collective, mehes2014collective, chate2008modeling, strombom2011collective, czirok1999collective, warren2018collective, zhang2010collective, papadopoulou2023dynamics, beppu2024magnetically}. These collective dynamics often emerge in various complex environments, such as in the presence of walls or confinement channels. In such conditions, it has been observed that the systems undergo directional motion instead of propelling in a chaotic manner \cite{direct_collective_motiom_goldstein, vincenti2019magnetotactic, doostmohammadi2019coherent, samui2021flow, khaluf2017scale, xu2019self}. Such dynamics emerge because of the hydrodynamic interaction with the obstacle. Researchers have performed various simulations \cite{joshi2023disks, ravnik2013confined, wen2023collective, negi2023geometry, doostmohammadi2019coherent, fazli2021active, chandragiri2019active, meng2018clustering} and experimental studies \cite{direct_collective_motiom_goldstein, doostmohammadi2019coherent, hardouin2020active} to unveil how such motion is developed and the kind of interaction that governs such collective motion\cite{keogh2024active, kempf2019active, hardouin2020active, fazli2021active, chandragiri2019active, meng2018clustering, thery2020self}. It has been observed from these studies that by changing the shape of the confinement, such as circular, annular, and one-dimensional channels, different kinds of swimming dynamics emerge, such as spiral vortex formation, dancing modes, or unidirectional motion\cite{joshi2023disks, samui2021flow}. 

Researchers have also tried to emulate similar collective states and directional motion in artificially engineered colloidal particles since they provide additional aspects for controlling the interaction and, hence, the dynamics \cite{ all_optical_propulsion_force, tailored_optical_propulsion, roundtrip_motion_janus, active_motion_of_janus_defocused, janus_trochoidal_dynamics, sile_nic_chormaic_janus_propulsion_nanofiber,thermotaxis_Cichos, engay2024transverse, wang2024nanoalignment, schmidt2023tunable, schmidt2019light, cui2024synchronization, mahdi2022light, andren2021microscopic, volpe2023roadmap, liu2024robotic, simmchen2016topographical, nsamela2023colloidal, wang2023determination}. In this context, colloidal dynamics have been studied in various hard and soft confinements. Lithographic boundaries and channels are used as hard confinement \cite{misiunas2019density}, and optical potential landscapes have been used for soft confinement \cite{ashkin1986observation, grier2003revolution, dholakia2008optical, volpe_optical_tweezer_and_manipulation, dholakia_initiating_revolution, thalhammer2011optical, han2015evanescent, han2016optically}. In most of the currently available studies, the motion of the colloids is directly enabled by the externally applied fields that drive the particles either by actively steering the external field \cite{braun2013optically,liu2024robotic, peng2020opto} or by creating static asymmetric force distribution \cite{padgett2011tweezers, optical_force_arising_from_phase_graients, tailored_optical_propulsion, roundtrip_motion_janus, active_motion_of_janus_defocused, janus_trochoidal_dynamics, sile_nic_chormaic_janus_propulsion_nanofiber, rubinsztein2016roadmap, maggi2015micromotors}, which often require complex feedback protocols. However, similar dynamical states can emerge under static externally applied fields facilitated by the mutual interaction between colloidal objects \cite{reactive_optical_matter, chand2023emergence, paul2022optothermal}. In this context, Yuval Yifat et al. recently demonstrated the directional motion of a hetero-dimer in a ring-shaped optical illumination. The resulting motion was governed by differences in the scattering efficiencies of the individual colloidal objects, inducing an imbalance \cite{reactive_optical_matter}. However, they reported a lack of control over the dynamics. Recently, we have shown that the optothermal interaction\cite{opto-thermophoretic_manipulation, optical_manipulation_heats_up, braun2013optically} between the colloids with different absorbing nature under focused Gaussian optical illumination results in directional rotation of the collective colloidal structures \cite{chand2023emergence}. In that case, the motion of the colloidal structures is localized to the laser beam spot about which the structures spin. Similar optothermal interaction-driven motion can be harnessed, avoiding the localization aspect by using extended optical traps \cite{roundtrip_motion_janus,sile_nic_chormaic_janus_propulsion_nanofiber} and holographic optical tweezers\cite{optical_force_arising_from_phase_graients, tailored_optical_propulsion, liu2024robotic, padgett2011holographic, jesacher2006holographic, maurer2007tailoring, maurer2011spatial}. Currently, there is an imperative to address how new dynamical states of colloids can emerge from optothermal interaction with such a structured optical field.

\begin{figure}[ht!]
    \centering
    \includegraphics[width = 450 pt]{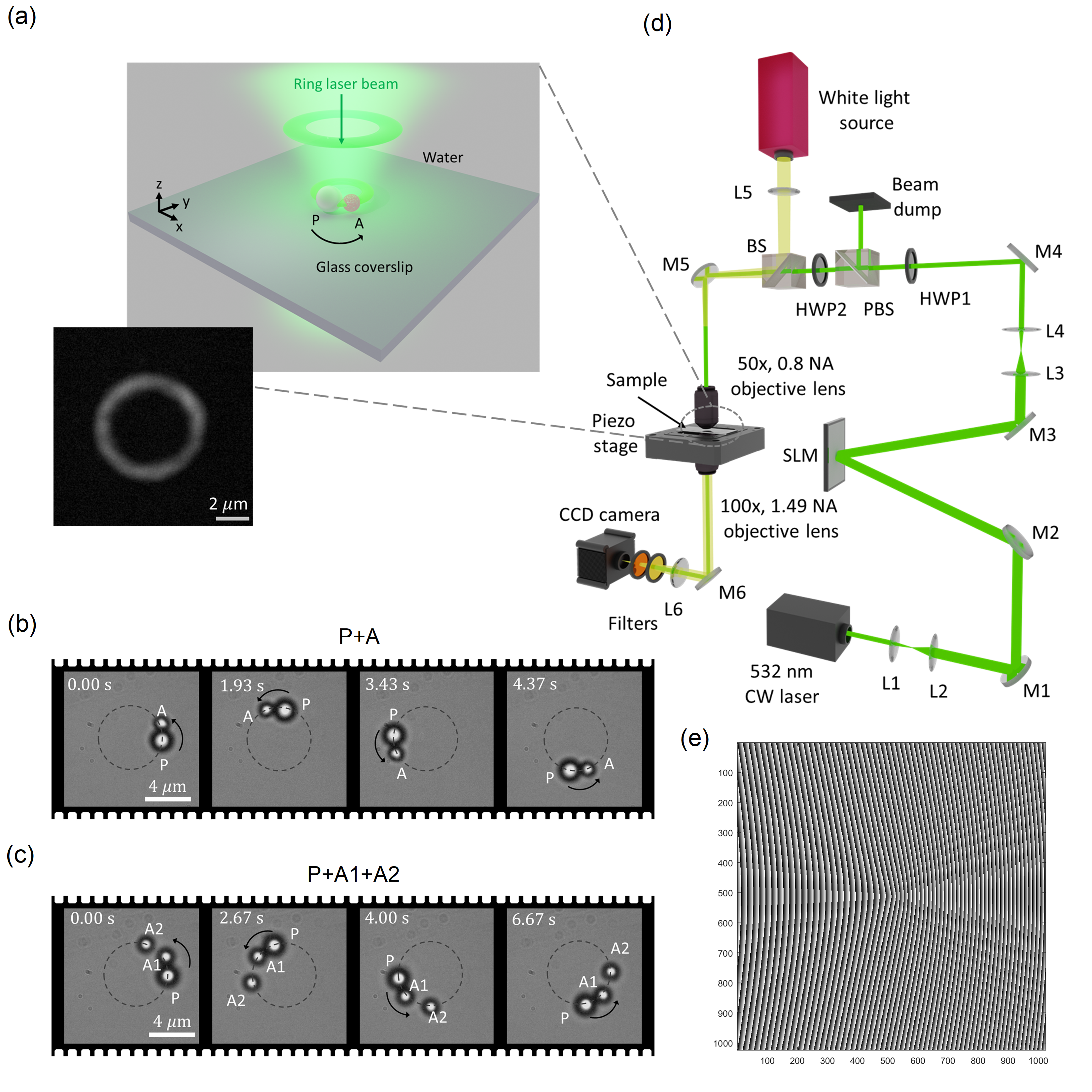}
    \caption{Dynamics of colloidal structures in a ring-shaped optical illumination. (a) The schematic of the experiment.  In a linearly polarized focused ring optical illumination of wavelength 532 nm laser beam, a colloidal dimer structure (composed of one passive (P) and one thermally active (A) colloid) exhibits directional revolution, where the direction is indicated by the black arrow. The experimental intensity profile is shown in the inset. The time series of a dimer (P+A) and trimer (P+A1+A2) structures are shown in (b) and (c), respectively. (d) A dual-channel optical experimental setup was used, where the ring optical field, generated by SLM, is coupled to the sample plane using a 50x, 0.8 NA objective lens. The signal is collected through a 100x, 1.49 NA oil objective lens and projected to a CCD camera to record the dynamics. (e) The corresponding blazed axicon hologram pattern was used to generate the ring optical field.}
    \label{Figure_1}
\end{figure}

Motivated by this, in this article, we show when colloids with different absorbing natures are confined in a ring-shaped optical illumination, they can undergo a directional revolution, where the optothermal interaction between the colloids fuels the motion on an optically laid out path. figure \ref{Figure_1}(a) shows the schematic of the experiment. The experimental intensity profile of the ring-shaped optical illumination used for the experiment is shown in the inset. The observed directional revolution for a dimer (composed of a passive (P) and a thermally active (A) colloid) and trimer structures (composed of a passive (P) and two thermally active colloids(A1, A2)) are shown in figure \ref{Figure_1}(b) and \ref{Figure_1}(c), respectively. Such directional revolution is absent when individual colloidal particles are considered. We also show control of revolving dynamics and interparticle distances between the particles in the colloidal structures by modulating the attractive and repulsive interaction using the intensity of the incident optical illumination. In this study, we use ring-shaped optical illumination since that reduces our dynamics to a pseudo-one-dimensional motion. Such illumination acts as a continuum field where colloids can exhibit continuous motion.

\section*{Materials and methods}
In our experiments, we use two types of colloids – passive and thermally active denoted by P and A, respectively. As passive colloids, we use melamine formaldehyde colloids of diameter 2 $\mu$m, and as thermally active colloids, iron oxide nanoparticles-infused polystyrene colloids of diameter 1.3 $\mu$m. Under optical illumination, the iron oxide nanoparticles in the thermally active colloids absorb laser energy and dissipate as heat, setting up a temperature field in the surrounding medium. Our previous research articles show the corresponding absorption spectra of the thermally active colloids \cite{ paul2022optothermal, chand2023emergence}. A dilute aqueous suspension of thermally active and passive colloids is prepared by mixing the two types of colloids in Milli-Q water. 10 $\mu$L of the mixture solution is enclosed in a sample chamber of height 120 $\mu$m between two glass coverslips. The sample cell is placed in a dual-channel optical microscopy setup, as shown in figure \ref{Figure_1}(d). A blazed axicon hologram pattern, as shown figure \ref{Figure_1}(e), produces the optical ring beam using a phase-only SLM from a continuous wave Gaussian laser source of wavelength 532 nm. The first-order diffracted optical ring beam from the SLM is collected using an iris and projected into the microscope objective lens (50X, 0.8 NA) by a relay lens combination (L3 and L4). This linearly polarized optical ring beam heats and traps the thermally active and passive colloids. To record the dynamics, the signal is collected from the sample plane using a 100X, 1.49 NA oil immersion objective lens and projected to a CCD camera (30 fps). The trajectories of the microparticles have been tracked from the recorded videos using Trackmate \cite{tinevez2017trackmate, schindelin2012fiji}.

\section*{Results and discussions}
\subsection*{Dynamics of colloids in an optical ring trap}
\begin{figure}[ht!]
    \centering
    \includegraphics[width = 350 pt]{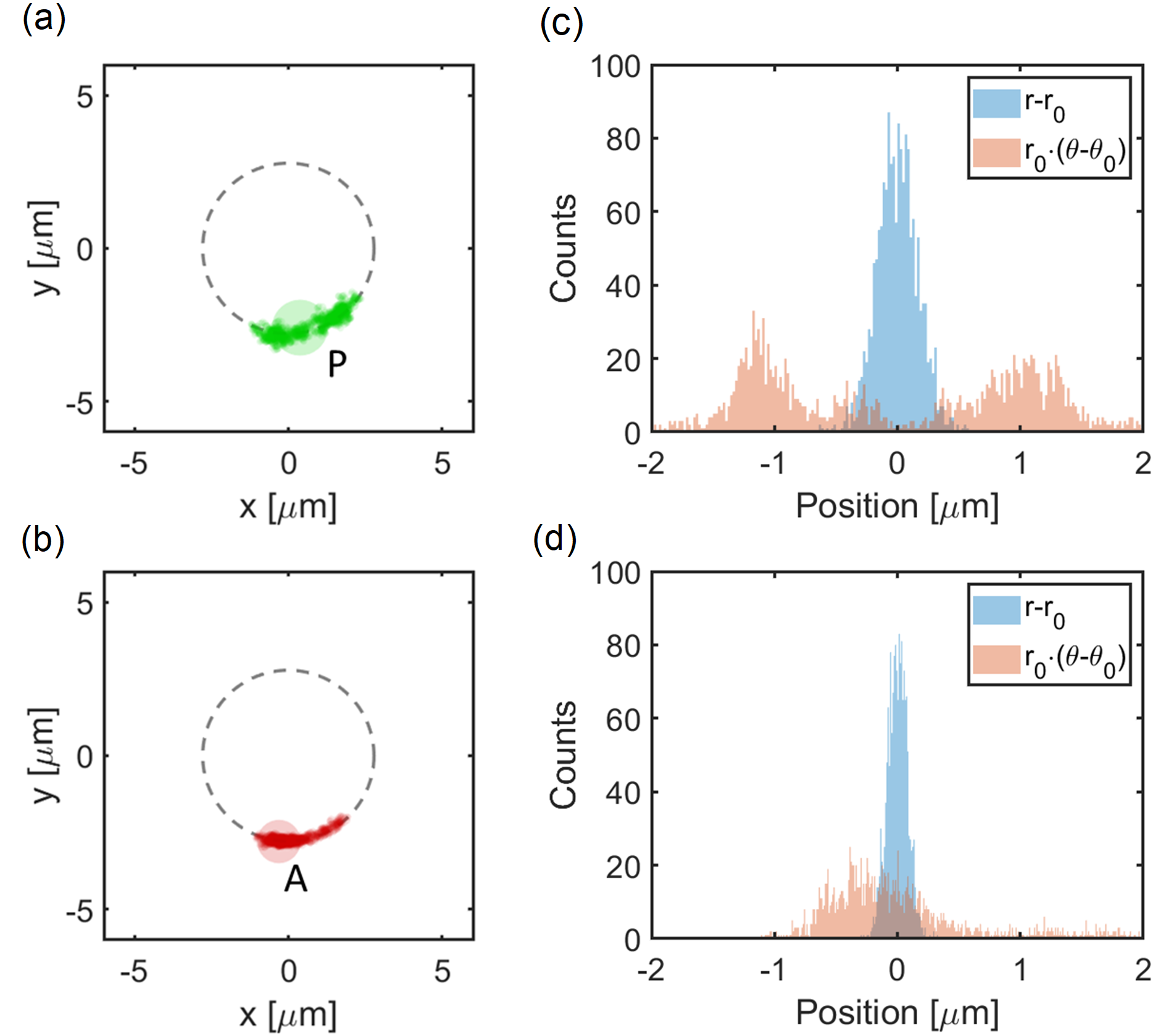}
    \caption{Thermally active and passive colloids under ring optical beam illumination with radius $r_0=2.79$ $\mu$m and intensity $I_0 =0.1$ mW/$\mu \mathrm{m}^2$. The two-dimensional position distribution of a passive colloid (P) of diameter 2 $\mu$m and a thermally active colloid (A) of diameter 1.3 $\mu$m are shown in (a) and (b), respectively. The trajectories are shown for 53s. In (c) and (d), the radial and angular position distributions for the colloids in (a) and (b) are shown.  For direct comparison of radial and angular confinement, instead of angular distribution ($\theta -\theta_0$), we have plotted the distribution of arc length ($ r_0 \cdot (\theta -\theta_0)$) along with the radial position distribution, where $\theta_0$ is the average angular position. From (c) and (d), it can be seen that the colloids P and A are confined radially, but azimuthally, they undergo diffusion.}
    \label{Figure_2}
\end{figure}

A ring-shaped optical intensity profile is generated in the sample plane as shown in the schematic figure \ref{Figure_1} (a) and described in the methods. The dynamics of passive and thermally active colloids under such optical illumination are shown by the two-dimensional trajectories in figure \ref{Figure_2}(a) and figure \ref{Figure_2}(b), respectively. From the trajectories, we can observe that when colloids A and P are individually considered, they are confined radially, but azimuthally they are not strictly bound. Sometimes, due to the nonuniform intensity of the optical illumination, colloids also show a slight confinement in the azimuthal direction at higher laser intensities. Figure \ref{Figure_2}(c) and \ref{Figure_2}(d) show the corresponding radial and azimuthal position distribution of colloid P and A. For direct comparison, we have plotted the distribution of the arc length $r_0(\theta-\theta_0)$ in figure \ref{Figure_2}(c) and \ref{Figure_2}(d) instead of the exact azimuthal position distribution ($\theta-\theta_0$). The radial confinement of the colloids is due to the gradient optical force acting on them, which traps the colloids near the maximum intensity region of the optical ring beam, i.e., at $r=r_0$ (see supporting information S1). 

\begin{figure}[ht!]
    \centering
    \includegraphics[width = 470 pt]{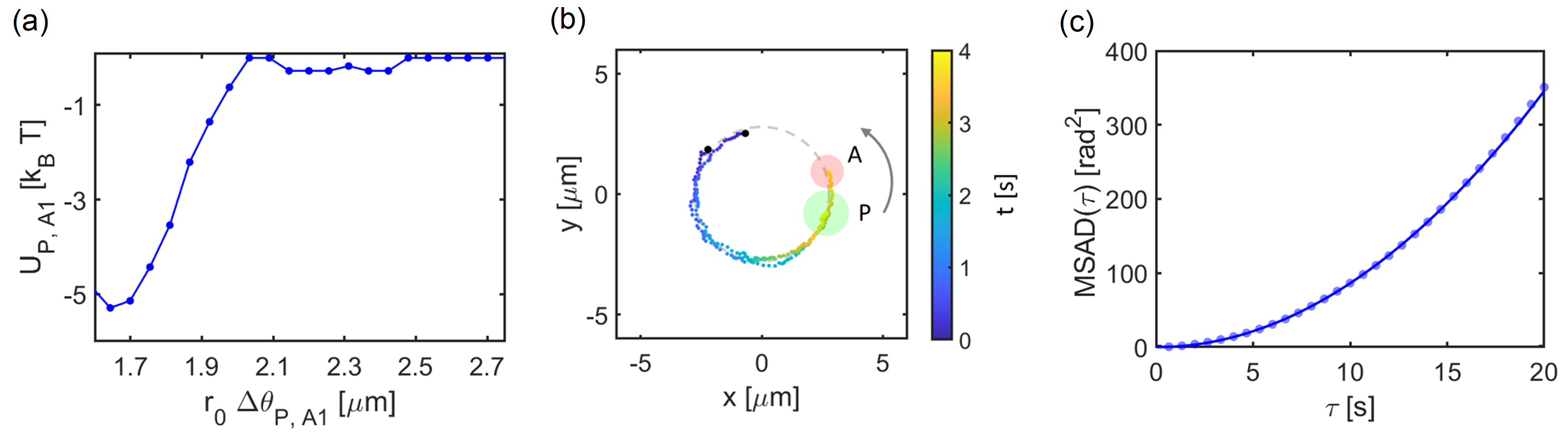}
    \caption{Colloidal dimer under ring optical beam illumination with radius $r_0=2.79$ $\mu$m and intensity $I_0 =0.5$ mW/$\mu \mathrm{m}^2$. When both the thermally active (A) and passive (P) colloids (of diameters 1.3 $\mu$m and 2 $\mu$m) are present under the ring beam, the thermal field produced by the thermally active colloid facilitates a unidirectional attractive interaction, which results in the revolution of the dimer structure. The corresponding interaction potential is shown in (a). (b) The resulting two-dimensional position distribution of the colloidal dimer. (c) The mean squared angular displacement of the dynamics.}
    \label{Figure_3}
\end{figure}

Although both the passive and thermally active colloids under the optical ring beam show qualitatively similar dynamics, the thermally active colloids absorb the laser energy and dissipate as heat, resulting in setting up a temperature distribution, which is quantified in supporting information S2. Being at an elevated temperature, these thermally active colloids induce an inward slip fluid flow which may drag other colloids towards the heat center (see supporting information S3). Thus, when both active and passive colloids are simultaneously present under the optical ring confinement, the slip flow-induced drag forces enable a unidirectional attractive interaction between the colloids. We can quantify this attractive feature by calculating the interaction potential ($ U_{A, P} $) between the thermally active and passive colloids\cite{Bechinger_3body_int_phys_rev_E, Bechinger_3body_int_PRL}. For that, at first, we obtained the distance distribution between the colloid A and P along the arc length $ P_{A, P} (r_0 \Delta \theta_{A, P})$, then we take the negative logarithm of this distribution to get the total interaction potential. 
 \begin{equation}
     U_{A, P} (r_0 \Delta \theta_{A, P}) \approx -\ln{P(r_0 \Delta \theta_{A, P})}
 \end{equation}
From the calculated  $U_{A, P}$, plotted in figure \ref{Figure_3}(a), it can be clearly noted that the potential minima is close to the minimum possible inter-particle distance between the colloids, i.e., $a_P+a_A = 1.65$ $\mu$m, which re-emphasizes attractive interaction between colloids A and P. Due to this unidirectional attractive interaction, the active and passive colloids form a dimer structure and collectively start revolving, as depicted in the two-dimensional trajectory in figure \ref{Figure_3}(b) (also see the SI video 1). To characterize the motion of this dimer, we have calculated the mean squared angular displacement ($MSAD (\tau)$), which is shown in figure \ref{Figure_3}(c). The experimental MSAD fits well with the equation,
\begin{equation}
    MSAD(\tau) = \langle{ [\theta (t+\tau)-\theta(t)]}^2\rangle = 2 D_r \tau +\omega ^2 \tau^2,
\end{equation}
where $D_r$ is the collective diffusion constant of the dimer in the azimuthal direction, $\omega$ is the angular velocity of rotation, and $\tau$ is the lag time. Since similar dynamics are not observed for the individual colloids and are only observed when both types of colloids are present simultaneously and facilitated by the optothermal interaction between them, we call such a dimer an active dimer (AD). We can obtain the average linear velocity of the colloidal dimer as $v= \omega \cdot r_0$.

\subsection*{Modeling using Brownian dynamics simulation}
To further understand how the collective effect of the thermo-osmotic flow-induced drag forces and optical gradient force leads to the experimentally observed dynamics, we have computationally simulated the dynamics of such active dimers. In our simulation model, we numerically solved Langevin's equation with the external forces \cite{self_phoretic_simulation_2022, volpe_active_simulation}. As external forces, we have considered (1) gradient optical forces and (2) thermo-osmotic slip flow-induced forces, as discussed below.

\begin{figure}[ht!]
    \centering
    \includegraphics[width = 350 pt]{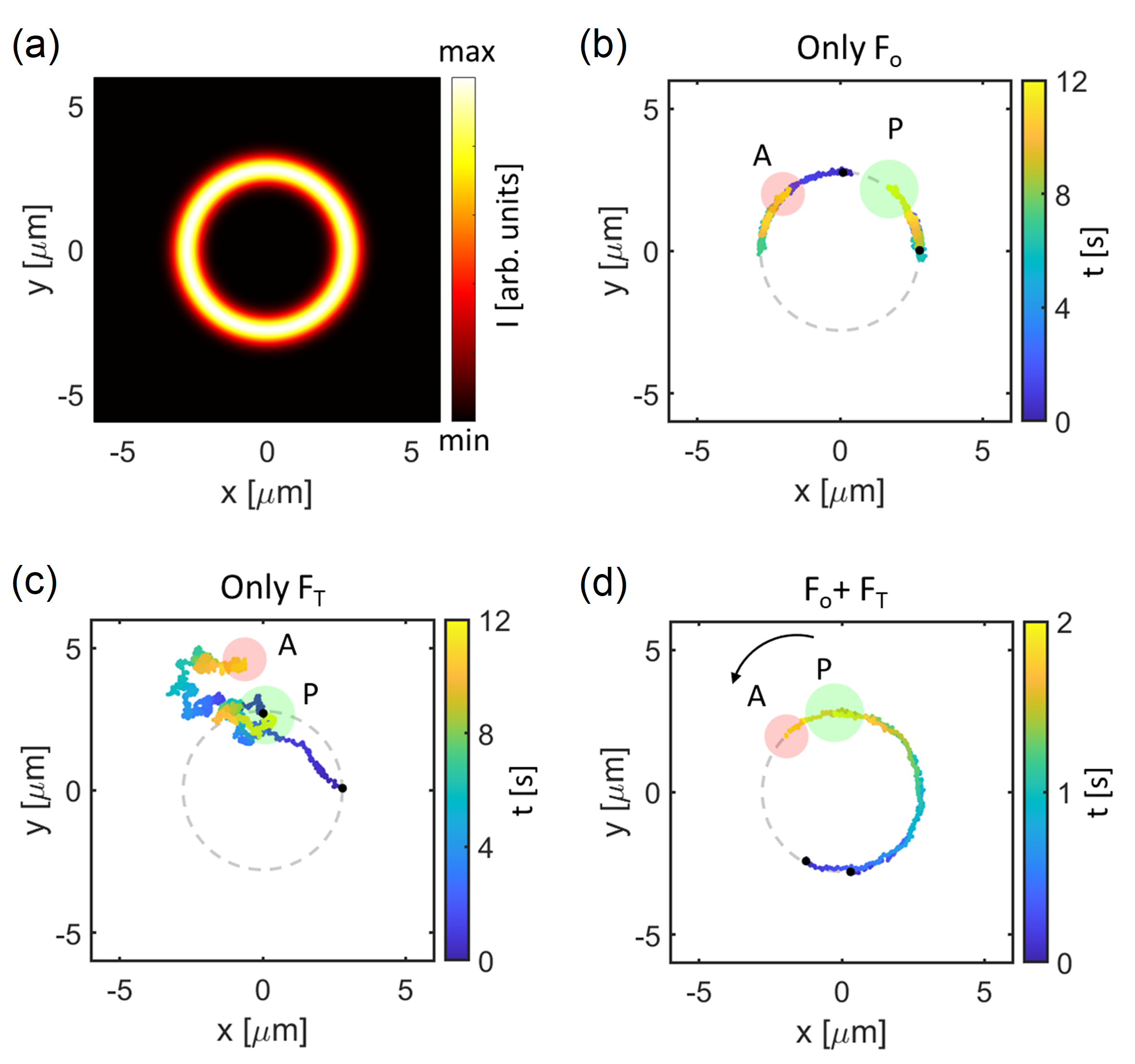}
    \caption{Simulation of dynamics of active dimers. The intensity profile of the ring optical beam profile, governed by the equation $I(r)= I_0  \exp{(-(r-r_0)^2/{w_0^2})}$, is shown in (a). In such an optical beam, the dynamics obtained by only considering the optical gradient force (${\Vec{F}}_o$) is shown in (b). (c) When only thermo-osmotic interaction is considered between the colloids, the directional dynamics are also not observed. (d) As both the optical and thermo-osmotic interactions are considered simultaneously, the assembled dimer structures lead to the directional revolving motion.}
    \label{Figure_4}
\end{figure}

The intensity distribution of an optical ring beam can be written as $$I(r)= I_0  \exp{(-(r-r_0)^2/{w_0^2})},$$ where $I_0$, 
 $r_0$ and $w_0$ are the maximum intensity of the distribution, the maximum intensity radius, and the beam waist of the optical ring beam. The corresponding intensity pattern governed by the equation is shown in figure \ref{Figure_4}(a), indicating that the simulated intensity field is qualitatively similar to the experimental ring beam intensity pattern. In such an optical beam, the gradient optical force ( $\Vec{F}_{o, i}$) acting on the colloids ($i=$ P, A) can be approximated as $\Vec{F}_{o, i} \simeq k_i (\Vec{r}_i - \Vec{r}_0)$, where this equation is valid for $ |\Vec{r}_i - \Vec{r}_0| << w_0$.

The thermo-osmotic drag force acting on P due to the thermo-osmotic slip flow ($v_s$) produced by the heated thermally active colloid A can be written as $ \Vec{F}_{T{\{P, A\}}} \approx \gamma_P \vec{v}_s \approx \sfrac{\gamma_P \chi \Vec{\nabla} T_A}{Ts} $, where $\Vec{\nabla}T_A$ is the temperature gradient produced by A at the location of P, and $ \gamma_P $ (see supporting information S4) and  $\chi$ are the viscous drag coefficient of colloid P and thermo-osmotic slip coefficient, respectively.
In our simulation, to ensure the exclusion region for one colloid due to the presence of other colloids, we considered the Lennard Jones interaction ($ \hat{\Vec{LJ}}_{\{ij\}}$) between the colloids which is written as, 
\begin{equation}
    \Vec{LJ}_{\{ij\}} =  4\epsilon { \bigl( 12 \frac{{\sigma_{ij}^{12}}}{{r_{ij}^{13}}} - 6 \frac{{\sigma_{ij}^{6}}}{{r_{ij}^{7}}}  \bigr)},
\end{equation}

where $ \sigma_{ij}=\sigma_0 (a_i+a_j) $ and $\epsilon$ are the Lennard Jones parameters, $\hat{r}_{ij}$ is the unit vector along the position vector of colloid $i$ with respect to colloid $j$,  $a_i$, and $a_j$ are the radius of $i^{th}$ and $j^{th}$ colloids ($i=$ P, A) respectively. In our simulation, we have used $\sigma_0 = 1$, $\epsilon = 20 \times 10^{-21}$ J.

With all the forces mentioned above, the Langevin's equation for colloidal pairs in AD can be written as,

\begin{equation}
   \gamma_A \Dot{\Vec{r}}_A (t) =  \Vec{F}_{o, A} (t) + \Vec{LJ}_{ \{A, P \} } (t)+ \sqrt{2 k_B T_A \gamma_A} \Vec{W}_A^t
\end{equation}

\begin{equation}
    \gamma_P \Dot{\Vec{r}}_P (t) =  \Vec{F}_{o, P} (t) + \Vec{LJ}_{\{P, A\} } (t)+ \Vec{F_T}_{\{P, A\}}(t) + \sqrt{2 k_B T_P \gamma_P} \Vec{W}_P^t
\end{equation}
The last term in both equations indicates the stochastic noise-induced random force on the colloids. To numerically solve, we need to put the above two coupled equations in discretized form. In discretized form  $ {\dot{\vec{r}}}^{(k)} = \bigl( {\vec{r}}^{(k+1)} - {\vec{r}}^{(k)} \bigr)/\Delta t$, and $ {\Vec{W}}^{(k)} = {\Vec{w}}^{(k)}/\sqrt{\Delta t}$, where $\Delta t$ is the time step, stochastic noise ${\Vec{w}}^{(k)}$ has Gaussian distribution with an average value of zero, and $(k)$ is indicating $k^{th}$ iteration. Thus, the above equations in discretized form can be written as,
\begin{equation}
    {\Vec{r}}_A ^{(k+1)} = {\Vec{r}}_A ^{(k)} + \frac{\Delta t}{\gamma_A} {\bigl( \Vec{F}_{o, A}^{(k)} + \Vec{LJ}_{ \{A, P \} } ^{(k)} + \sqrt{\frac{2 k_B T_A \gamma_A}{\Delta t}} \Vec{w}_A^{(k)} \bigr)}
\end{equation}

\begin{equation}
    {\Vec{r}}_P ^{(k+1)} = {\Vec{r}}_P ^{(k)} + \frac{\Delta t}{\gamma_P} {\bigl( \Vec{F}_{o, P}^{(k)} + \Vec{LJ}_{ \{P, A \} } ^{(k)} +  \Vec{F_T}_{\{P, A\}}^{(k)} +\sqrt{\frac{2 k_B T_P \gamma_p}{\Delta t}} \Vec{w}_P^{(k)} \bigr)}  
\end{equation}

By numerically solving these equations, we can obtain the dynamics of the colloidal dimer. We can understand the role of both these forces by considering them nonzero individually and solving the Langevin equation. At first, when we considered only nonzero optical gradient force acting on the colloids. Then, the colloids are found to be only confined radially, but no revolving motion is observed, as can be seen from figure \ref{Figure_4}(b). When we consider non-zero thermo-osmotic force only, due to stochastic noise, the colloids move out of the optical illumination region in the absence of any confinement force. As the colloids are away from the optically illuminated region, they are not heated up, and they exhibit diffusive behavior, as can be seen from figure \ref{Figure_4}(c). In contrast, if both the optical gradient force and the slip flow-induced drag force are considered simultaneously, the dimer structure exhibits directional revolution, as shown in figure \ref{Figure_4}(d). The slip flow-induced drag force leads to directional motion from passive to thermally active colloids, and the gradient optical force confines motion to the optically laid out path in a ring beam.

\subsection*{Control over the revolution}
\begin{figure}[ht!]
    \centering
    \includegraphics[width = 350 pt]{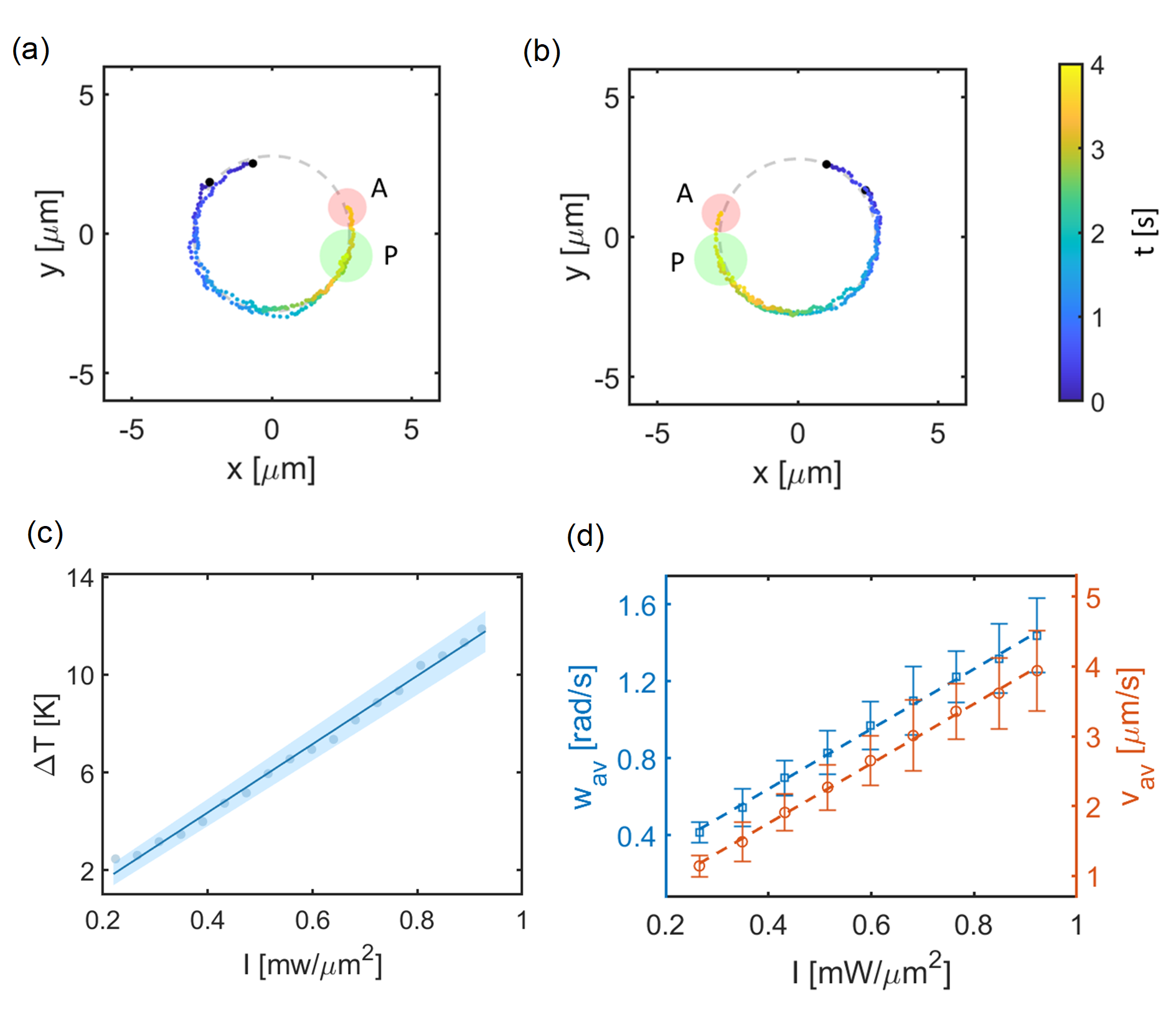}
    \caption{Control over the dynamics. The direction revolution of the AD structure depends on the arrangements of the constituent colloids in the structure, and the active dimer propels heading towards the thermally active colloid in the structure. The trajectories of two such oppositely revolving dimer structures, based on the arrangement of the constituent colloids, have been shown in (a) and (b). (c) The linear incremental trend of the surface temperature of the thermally active colloid with the intensity of incident ring optical illumination has been shown. (e) Both angular $\omega$ and linear velocities $v$ of revolution indicate a linear incremental trend (dashed lines) with the intensity of the incident optical field, which can be attributed to the increase in the thermal field.}
    \label{Figure_5}
\end{figure}

Since the interaction between the colloids in the active dimer is attractive and unidirectional, acting from passive to thermally active colloids, changing the arrangement of the colloids, we can control their direction of revolution (see SI video 2). figure \ref{Figure_5}(a) and \ref{Figure_5}(b) show the trajectories of two ADs revolving in opposite directions based on their arrangements. Not only the direction, we can also control the velocity by modulating the induced temperature field produced by the thermal active colloids. By tuning the intensity of the optical field, we can control the thermal field and, hence, the speed of the dynamics (see SI video 3). Figure \ref{Figure_5}(c) shows the linear incremental trend of the surface temperature of thermally active colloids with ring optical field intensity. As a result, both the angular speed ($\omega$) and the linear speed ($v$) increase linearly with the intensity of the incident optical beam, as shown in figure \ref{Figure_5}(d).

\subsection*{Simultaneous attractive and repulsive interaction in active trimer structures}

\begin{figure}[ht!]
    \centering
    \includegraphics[width = 480 pt]{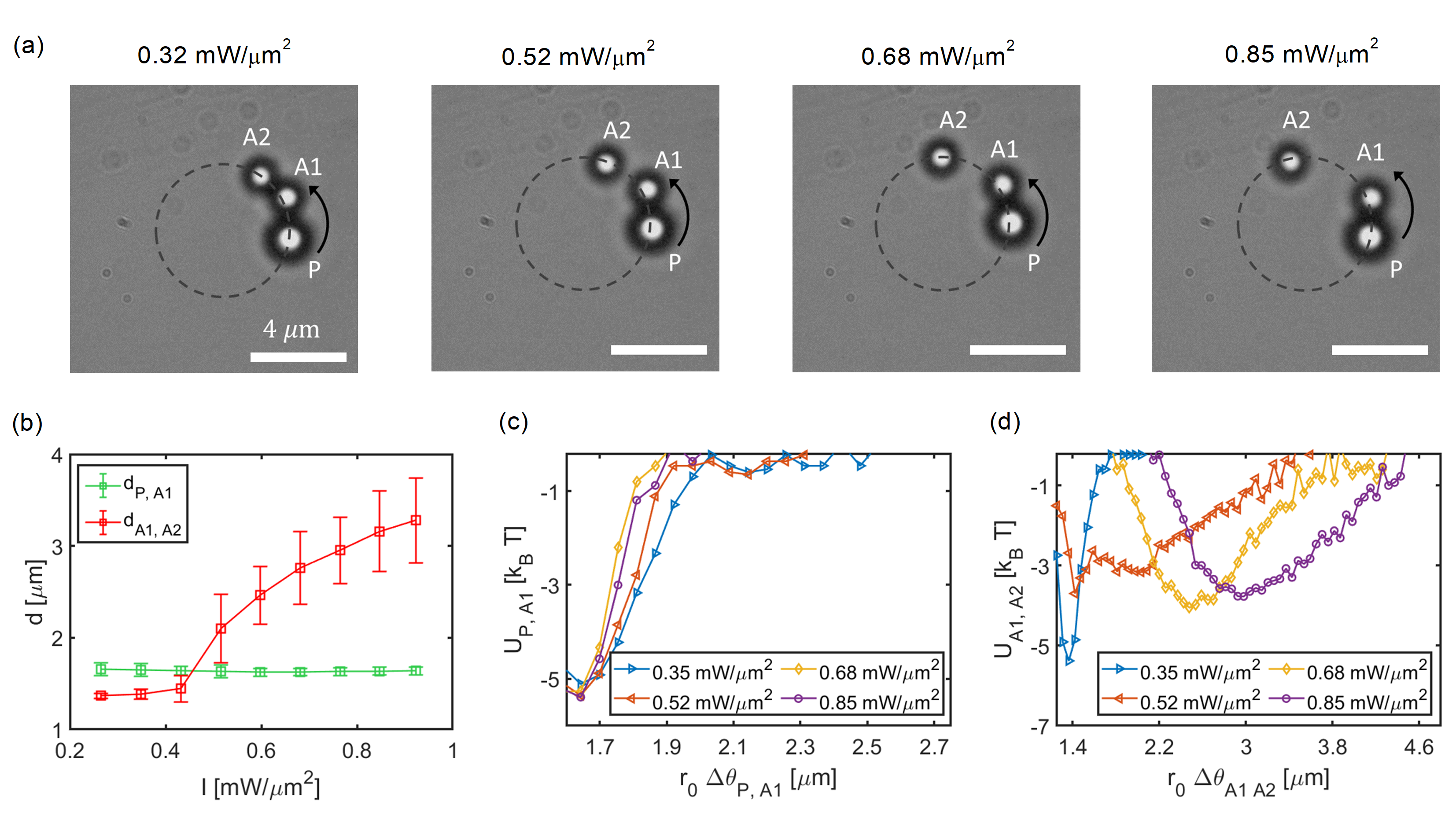}
    \caption{Tuning the interactions between the colloids in the colloidal active trimer (AT) structures. (a) The interaction between the colloids forming the trimer structure and the governing dynamics can be tuned by the intensity of the incident optical illumination depending on the interplay between the thermo-osmotic flow-induced attractive and thermophoretic repulsive interaction. (b) The corresponding distances between colloids facilitated by such interaction as a function of laser intensities show that the distance between the colloidal pair (P, A1) does not change significantly, whereas the distance between the colloidal pair (A1, A2) exhibits an incremental trend. The calculated interaction potential between the colloids P, A1, and A1, A2 at different laser intensities have been shown in (c) and (d).}
    \label{Figure_6}
\end{figure}

Up to the previous section, we have seen the emergence of directional revolution facilitated by the attractive thermo-osmotic flow-induced interaction between the colloids in the active dimer structure. In this section, we will discuss the simultaneously attractive and passive interaction between colloids forming an active trimer (AT) structure, which is composed of one passive (P) and two thermally active (A1, A2) colloids. These active trimer structures undergo similar directional revolutions under optical ring beam confinement, as seen in the case of active dimer structures. But in contrast to the dimer structure here, the intensity of the optical field not only modulates the angular velocities but their interparticle distances can also be modulated, as shown in figure \ref{Figure_6}(a) (see SI video 4). From figure \ref{Figure_6}(b), we can see the distance between the colloidal pair (P, A1) does not change significantly, whereas the distance between A1 and A2 changes significantly with the intensity of the incident optical field. To understand this phenomenon, we have the corresponding interaction potentials ($U_{P, A1}$, $U_{A1, A2}$) between P, A1, and A1, A2  have been calculated and plotted in figure \ref{Figure_6}(c) and \ref{Figure_6}(d) respectively. We can clearly observe the shift in the potential $U_{A1, A2}$ minima position, which is not observed for $U_{P, A1}$, which suggests that in addition to the attractive slip flow-induced drag force, there is laser intensity-dependent repulsive interaction developed between the thermally active colloids. Similar repulsive interaction between thermally active colloids has already been reported \cite{paul2022optothermal}. Such repulsive interaction can originate due to the thermophoretic interaction between the colloids. The thermophoretic interaction is characterized by the thermo-diffusion constant ($D_T$) of colloids, and the temperature dependence of the thermodiffusion constant can be written as $D_T(T) \approx C(T^*-T)$, where C is a constant, $T^*$ is the transition temperature \cite{iacopini_DT_dependence, helden_direct_measurement_of_thermophoretic_forces, piazza_thermophoretic}. For $T<T^*$, $D_T(T)$ is negative, which suggests the colloids are attracted towards the high-temperature regime, whereas $T>T^*$, $D_T(T)$ is positive, suggesting the colloids are to be repelled from the high-temperature regime. Due to this temperature-dependent thermophoretic behavior, a repulsive interaction is developed between the thermally active colloids above a certain laser intensity. In contrast, we did not observe any repulsive interaction between the colloidal pair P and A1. This might be because of two reasons: (1) the passive colloids do not get directly heated up under laser illumination, and no significant change of thermodiffusion constant has occurred. (2) The thermophoretic transition temperature $T^*$ for the passive colloid is beyond the experimentally probed temperature increment.

\begin{figure}[ht]
\centering
    \includegraphics[width = 485 pt]{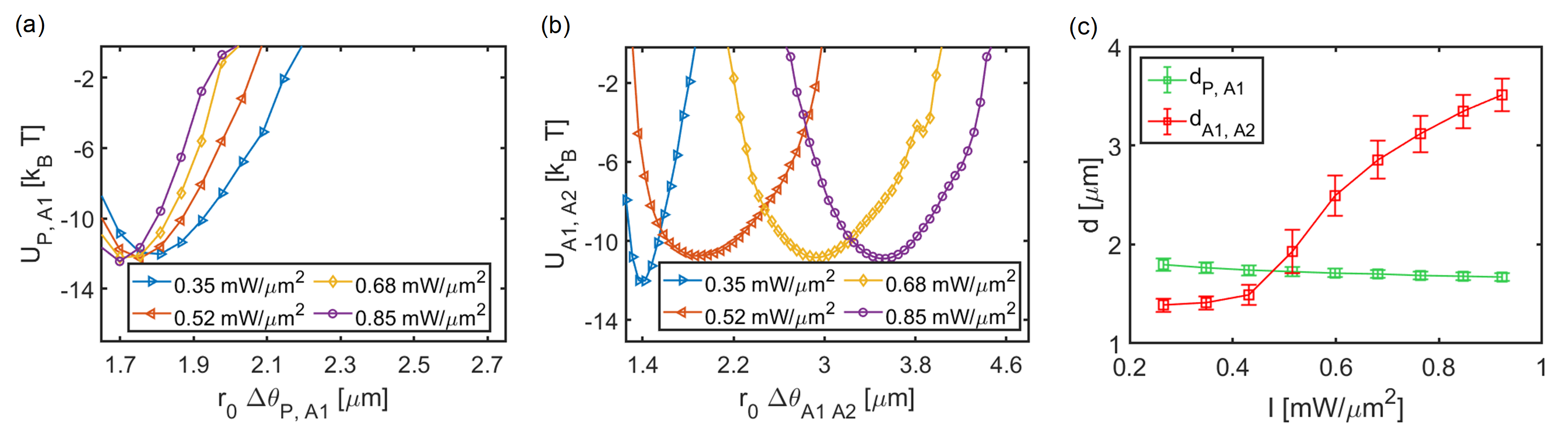}
    \caption{ Simulation of interaction between colloids in AT structure with one passive (P) and two thermally active (A1 and A2) colloids. The interaction potentials $U_{P, A1}$ and $U_{A1, A2}$ between the colloidal pairs (P, A1) and (A1, A2) are shown in (a) and (b), respectively. The position of potential minima for $U_{P, A1}$ does not change significantly with incident laser intensity, whereas for $U_{A1, A2}$, the minima position is shifted with the intensity of the optical illumination. (c) The simulated average distance between the colloidal pairs (P, A1) and (A1, A2).}
    \label{Figure_7}
    
\end{figure}

To validate our experimental understanding of the dynamics of active trimers, we have also simulated the dynamics of three similar particle-level colloidal systems under optical ring illumination. The simulation details are given in the supporting information S5. The simulated interaction potentials between the colloidal pair (P, A1) and (A1, A2) are shown in figure \ref{Figure_7}(a) and figure \ref{Figure_7}(b), respectively. From the simulation, we also clearly observe a similar shift in the interaction potential $U_{A1, A2}$ minima with incident optical illumination, and no significant shift is observed in the case of $U_{P, A1}$. The simulated distances between the colloids are shown in figure \ref{Figure_7}(c), which also exhibit a similar trend as figure \ref{Figure_6}(b). 

\section*{Conclusions}
In this work, we studied the dynamics of colloidal structures (dimers and trimers) due to the optothermal interaction between colloidal objects under ring-shaped optical illumination. Both attractive and repulsive interactions emerge due to the interplay between the thermo-osmotic and thermophoretic interactions. The thermo-osmotic slip flow-induced drag force enables the unidirectional attractive interaction that leads to a directional revolution in the optically laid-out path. The corresponding direction of revolving structures is dependent on the relative arrangement of the constituents' passive and thermally active colloids. Meanwhile, the speed of the motion can be controlled by the intensity of the incident optical field. We have also observed repulsive interaction between the heated thermally active colloids above certain laser intensity where the positive thermophoretic effects dominate over the slip-flow induced attractive interaction between them. Due to the interplay between these attractive and repulsive interactions, we can also modulate the distances between the colloids without altering the direction of motion. Similar motion can be adapted in various other shapes of optical illumination. In particular, we have also observed similar dynamics with line optical profiles, which suggests the generalizability of these interactions and the governing dynamics. Such directional motion with tunable distances between the colloids can be very useful for various biomedical and technological applications, such as transporting cargo and medication to infected cells and microsurgeries \cite{andren2021microscopic, friese2001optically}. This can potentially be very useful for understanding similar collective motion for various natural systems in the presence of walls, obstacles, and channels \cite{direct_collective_motiom_goldstein, vincenti2019magnetotactic, doostmohammadi2019coherent}. Besides these, since our experimentally obtained dynamics are governed by the thermo-osmotic and thermophoretic interaction between the colloids, both these interactions are reported to be of thermodynamic origin \cite{derjaguin1987surface, hydrodynamic_manipulation_cichos}. Thus this study can potentially be very useful as a model system for understanding out-of-equilibrium thermodynamics at the microscale.

\section*{Acknowledgement}
R.C. and A.S. thank Diptabrata Paul, Chaudhary Eksha Rani, Sumant Pandey, and Sneha Boby for the fruitful discussion and valuable input. R.C. thanks Ratimanasee Sahu and Prof. Vijayakumar Chikkadi for some valuable discussions. R.C. also thanks Prof. Clemens Bechinger for suggesting some essential research articles. R.C. thanks Prof. Umakant Rapol, Sudhir Lone, and Bhagyashri Kanade for helping us characterize our colloidal particles. 
\section*{Funding sources}
This work was partially funded by AOARD (grant number FA2386-23-1-4054) and the Swarnajayanti fellowship grant (DST/SJF/PSA-02/2017-18) to G.V.P.K.

\section*{Supporting information}
All the supporting informations of this article can be found in \url{https://drive.google.com/drive/folders/1RQSpgUtmQZaj5sH0YTM1SCHHPUVbNfKh?usp=sharing}.

\bibliographystyle{unsrt}
\bibliography{References_SI.bib}
\end{document}